%% file: main.tex
\documentclass[sigconf,letterpaper,10pt]{acmart}

\usepackage{url}
\usepackage[utf8]{inputenc}
\usepackage{xcolor}
\usepackage{amsmath}
\usepackage{xspace}
\usepackage{epsfig}
\usepackage{balance}
\usepackage{booktabs}
\usepackage{tabularx}
\usepackage[font=footnotesize]{subcaption}
\usepackage[font=footnotesize]{caption}
\usepackage{mathtools}
\usepackage{soul}
\usepackage{lipsum}
\usepackage{bm}
\usepackage{enumerate}
\usepackage{natbib}
\usepackage{enumitem}

\usepackage[acronyms,nonumberlist,nopostdot,nomain,nogroupskip,acronymlists={hidden}]{glossaries}
\newglossary[algh]{hidden}{acrh}{acnh}{Hidden Acronyms}
\glsdisablehyper

\usepackage{tikz}
\usepackage{pgfplots}
\pgfplotsset{compat=newest}
\pgfplotsset{plot coordinates/math parser=false}
\newlength\fheight
\newlength\fwidth
\usetikzlibrary{plotmarks,patterns,decorations.pathreplacing,backgrounds,calc,arrows,arrows.meta,spy,matrix,scopes}
\usepgfplotslibrary{patchplots,groupplots}
\usepackage{tikzscale}

\input{acronyms.tex}

\usepackage{algorithm}
\usepackage{algorithmicx}
\usepackage{algcompatible}
\usepackage[noend]{algpseudocode}
\usepackage{booktabs}
\usepackage{soul}

\title{Twinning Commercial Radio Waveforms in the Colosseum Wireless Network Emulator}

\author[D. Villa, D. Uvaydov, L. Bonati, P. Johari, J.M. Jornet, T. Melodia]{Davide Villa, Daniel Uvaydov, Leonardo Bonati, Pedram Johari,\\Josep Miquel Jornet, Tommaso Melodia}
\affiliation{%
  \institution{Institute for the Wireless Internet of Things, Northeastern University, Boston, MA, U.S.A.}
  \city{}
  \state{}
  \country{}
  }
\email{{villa.d, uvaydov.d, l.bonati, p.johari, j.jornet, melodia}@northeastern.edu}

\begin{abstract}

Because of the ever-growing amount of wireless consumers, spectrum-sharing techniques have been increasingly common in the wireless ecosystem, with the main goal of avoiding harmful interference to coexisting communication systems.
This is even more important when considering systems, such as nautical and aerial fleet radars, in which incumbent radios operate mission-critical communication links.
To study, develop, and validate these solutions, adequate platforms, such as the Colosseum wireless network emulator,
are key as they enable experimentation with spectrum-sharing heterogeneous radio technologies in controlled environments.
In this work, we demonstrate how Colosseum can be used to twin commercial radio waveforms to evaluate the coexistence of such technologies in complex wireless propagation environments.
To this aim, we create a high-fidelity spectrum-sharing scenario on Colosseum
to evaluate the impact of twinned commercial radar waveforms on a cellular network operating in the CBRS band.
Then, we leverage \acrshort{iq} samples collected on the testbed to train a machine learning agent that runs at the base station to detect the presence of incumbent radar transmissions and vacate the bandwidth to avoid causing them harmful interference.
Our results show an average detection accuracy of 88\%, with accuracy above 90\% in \acrshort{snr} regimes above $0$\:dB and \acrshort{sinr} regimes above $-20$\:dB, and with an average detection time of $137$\:ms.
\end{abstract}



\begin{document}


\maketitle

\glsresetall
\glsunset{cast}
\glsunset{usrp}
\glsunset{fpga}
\glsunset{uhd}



\section{Introduction}

The evolution of wireless technology has resulted in the ever-increasing complexity of wireless systems design. New generations of wireless networks have become more challenging to manage due to requirements necessitating optimal sharing of valuable resources between expanding sets of users. These challenges force researchers to think beyond traditional model-based approaches that are often limited to a specific problem scope and move toward
\gls{ai} solutions that offer superior performance in a wider range of conditions (e.g., channel conditions in this context) thanks to their data-driven nature.

The advancements of \gls{ai}-based solutions pave the way for more efficient use of the limited \gls{rf} spectrum, enabling multiple wireless systems to coexist harmoniously and cater to the growing demands of the digital era. Besides these benefits, challenges still exist regarding the potential interference between wireless networks that share the same spectrum. These mostly concern the adverse impact of such networks on incumbent radios with critical safety communication links~\cite{guiyang2022artificial}.
%
%
One example is the potential interference of 5G \glspl{ran}
on the incumbent radar communications in the \gls{rf} band ranging between $3.55$\:GHz and $3.7$\:GHz. This necessitates thorough study, research, and development of mitigation strategies to ensure the reliable operation of both systems, i.e., seamless communication of cellular networks while avoiding harmful interference to incumbent radar systems within nautical and aerial fleets~\cite{caromi2018detection}.
%
%
In the context of Open \gls{ran}, the \gls{ai}/\gls{ml} agents can be deployed in the \glspl{ric} proposed by O-RAN~\cite{polese2023understanding}, i.e., as xApps and rApps, or at the \gls{bs} directly as dApps~\cite{doro2022dapps}.
A use case of interest with such applications is to optimize the spectrum utilization and mitigate interference between coexisting wireless systems such as next-generation cellular, incumbent, or unlicensed radios~\cite{Baldesi2022Charm}. However, these methods require an abundance of high-quality data to train effective \gls{ml} models making them applicable to limited case studies that are often not scalable.
Gathering diverse and representative datasets that capture real-world scenarios can be time-consuming and resource-intensive, but is crucial for achieving accurate and robust \gls{ai} algorithm performance. As an alternative, high-fidelity emulation-based platforms can provide similar-quality data while offering several benefits compared to experimental setups: they are cost-effective, time-efficient, reproducible, and readily scalable~\cite{bonati2021colosseum}.
%
In the context of incumbent radios and communication links, such as nautical and aerial fleets, safety is of paramount importance, and non-carefully planned real-world experiments can endanger critical operations. Emulation platforms allow researchers to simulate interference scenarios and evaluate their impact without posing any actual risk to operational systems. However, collecting high-fidelity data not only requires an emulation platform that replicates the real wireless environment, but it also needs high-precision replicas of wireless nodes (core network, \gls{bs}, and \gls{ue}) that can reliably reproduce the network protocols and wireless waveforms to represent what happens in real-world deployments---a true wireless digital twin~\cite{villa2023dt}. 
Prior works, however, generate datasets that are not always able to capture high-fidelity and diverse environments, or that leverage synthetic waveforms that are not representative of commercial radios. This may result in impractical \gls{ai} models, whose performance substantially degrades when deployed in the real-world~\cite{gao2019deep, wang2019data, zheng2020spectrum}.

In this work, we develop a framework to emulate a spectrum-sharing scenario with cellular and radar nodes implemented in a high-fidelity digital twin system that can be reliably used to collect data, train \gls{ai} networks, and test them in realistic scenarios. We implement this framework on Colosseum, the world's largest wireless network emulator with hardware-in-the-loop~\cite{bonati2021colosseum}. We collect \gls{iq} samples of radar and cellular communications and train a \gls{cnn} that can be deployed as a dApp to detect the presence of radar signals and notify the \gls{ran} to stop operations to eliminate the interference on the incumbent radar communications. Our experimental results show an average accuracy of 88\%, with accuracy above 90\% in \gls{snr} regimes above $0$\:dB and \gls{sinr} regimes above $-20$\:dB. Through timing experiments, we also experience an average detection time of $137$\:ms.
This demonstrates the effectiveness of our system in detecting in-band interference in the \gls{cbrs} band, as it complies both with maximum timing requirements ($60$\:s) and accuracy (99\% within the $60$\:s time window)~\cite{goldreich2016requirements}.
%


The remaining of this paper is organized as follows. Section~\ref{sec:colosseum} overviews the Colosseum wireless network emulator.
Section~\ref{sec:radarusecase} details the integration of commercial radar technologies on Colosseum, as well as a spectrum-sharing scenario for the coexistence of cellular and radar signals.
Section~\ref{sec:radardetection} details our intelligent radar detection use case, while Section~\ref{sec:results} discusses our results.
%
Section~\ref{sec:conclusion} concludes the paper.


\section{Primer on Colosseum}
\label{sec:colosseum}

Colosseum is the world's largest wireless network emulator with hardware-in-the-loop hosted at Northeastern University and part of the \gls{pawr} Project~\cite{bonati2021colosseum, pawr}.
It consists of 256 NI/Ettus USRP X310 \glspl{sdr}, each equipped with two UBX-160 daughterboards able to irradiate signals between $10$\:MHz and $6$\:GHz.
Half of the \glspl{sdr} are allocated to the users and remotely accessible to carry out experiments on the system. This is done through the use of so-called \glspl{srn}---a combination of a high-compute server (48-core Intel Xeon E5-2650 CPUs with $126$\:GB of RAM) and an \gls{sdr} interfaced through a $10$\:Gbps connection---that the users of the system can reserve and program through softwarized \glspl{lxc}.
The other half is allocated to the \gls{mchem}, which is in charge of truthfully reproducing the conditions of heterogeneous wireless environments that the users can leverage in their \gls{sdr}-based experiments.
This is done through \gls{fir} filters implemented through an array of 64 Virtex-7 690T FPGAs.
When users transmit with Colosseum \glspl{srn}, the \gls{rf} waveforms generated by the \glspl{sdr} do not travel over the air but are sent to the \gls{mchem} \glspl{sdr} via coaxial cables.
This component, then, leverages the above-mentioned \gls{fir} filters to apply the channel conditions---expressed as a series of channel taps---to the signals generated by the users, and then transmit the signals resulting from this processing operation to the other \glspl{srn}.
Practically, channel taps are organized in a series of \gls{rf} scenarios that the users can choose from when performing their experiments.
Scenarios and taps are computed beforehand, e.g., through on-site measurements or software-based like ray-tracers~\cite{tehrani2021creating,villa2023dt}, installed on the Colosseum system, and made publicly available to the users.
Through them, users can prototype and evaluate different protocol stacks and waveforms with different channel conditions and node mobility---including effects such as fading, path loss, shadowing, different speeds, and movement trajectories---as if the radios were transmitting in the real-world environment that is emulated on the testbed.

\begin{figure}[ht]
    \vspace{-5pt}
    \centering
    \includegraphics[width=\columnwidth]{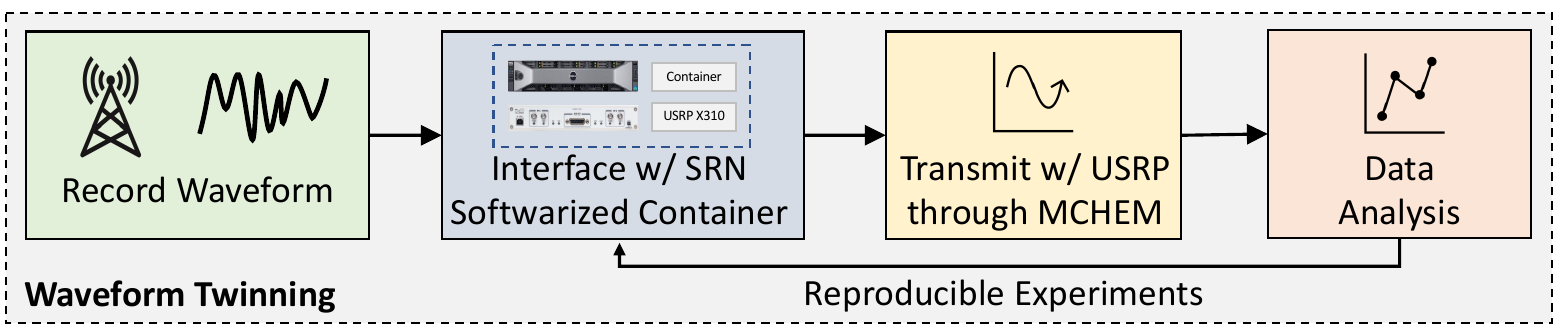}
    \caption{Waveform twinning on Colosseum.}
    \label{fig:waveformblock}
    \vspace{-5pt}
\end{figure}

\textbf{Waveform Twinning.}
Through the use of software containers, users can twin waveforms on the Colosseum \glspl{srn}. This process is shown at a high-level in Figure~\ref{fig:waveformblock}. First, the waveform is either recorded from a real-world transmission (e.g., radar, Wi-Fi, or cellular transmission) or synthetically generated.
The waveform is imported on Colosseum and interfaced with the softwarized \gls{lxc} container running on Colosseum \gls{srn}.
It is then transmitted by the \gls{srn} USRP to the other nodes of the experiment through \gls{mchem}.
At the end of the experiment, data is collected and analyzed for post-processing purposes.
Finally, it is worth noticing this procedure can be repeated on Colosseum for the fine-tuning of designed user solutions and their validation, thus allowing reproducible experiments to be carried out on the testbed.

\section{Coexistence of Cellular and Radar Technologies in CBRS Band}\label{sec:radarusecase}

In this section, we consider the use case of 4G/5G \glspl{ran} transmitting in the \gls{cbrs} band that needs to vacate said bandwidth because of an incoming radar transmission.
%
\gls{cbrs} regulations allow commercial broadband access to the \gls{rf} spectrum ranging from $3.55$\:GHz to $3.7$\:GHz, as depicted in the \gls{cfr}~\cite{cfr2016}. This spectrum is shared with various incumbents, including the U.S. military, which operates radar systems in this frequency range, e.g., shipborne radars along the U.S. coasts. According to the regulations, dynamic access to the spectrum is permitted as long as the network is able to detect the presence of the radar and activates interference mitigation measures when necessary~\cite{caromi2018detection}.
As it will be described in Section~\ref{sec:radardetection}, \glspl{bs} leverage \gls{ai}/\gls{ml} agents ---that can run as dApps--- to perform inference on the received \gls{iq} samples and detect incoming radar transmissions, which will be described in Section~\ref{sec:radar}.
Once detected, \glspl{bs} can either move to an unused bandwidth, if any, or terminate any ongoing communication to give priority to the radar.
To effectively study this use case in the Colosseum wireless network emulator, we developed an \gls{rf} propagation environment---located in the Waikiki Beach in Honolulu, Hawaii, described in Section~\ref{sec:rf-scenario}---in which a coastline \gls{bs} working in the \gls{cbrs} bandwidth needs to vacate said bandwidth due to the start of radar transmissions from a boat moving in the North Pacific Ocean.

\subsection{Radar Characterization}
\label{sec:radar}

Radar systems leverage reflections of \gls{rf} electromagnetic signals from a target to infer information on such target~\cite{mahafza2017introduction}.
Typical information may include detection, tracking, localization, recognition, and composition of the target, which may include aircraft, ships, spacecraft, vehicles, astronomical bodies, animals, and weather phenomena.
Even though radar's primary uses were mainly related to military applications, nowadays this technology is commonly used in other areas, such as weather forecasting, and automotive applications.

%

\begin{figure}[h]
    \vspace{-5pt}
    \centering
    \begin{subfigure}[b]{0.49\columnwidth}
        \includegraphics[width=0.99\columnwidth]{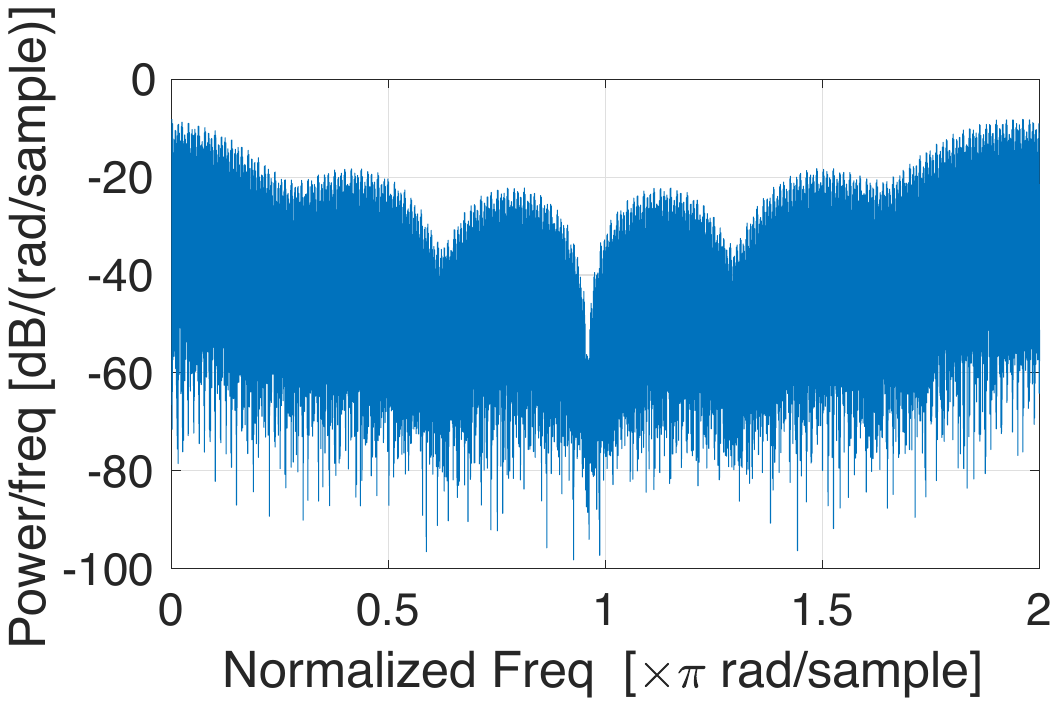}
        \caption{Radar \acrshort{psd}}
        \label{fig:radarpsd}
    \end{subfigure}
    \begin{subfigure}[b]{0.48\columnwidth}
        \includegraphics[width=0.99\linewidth]{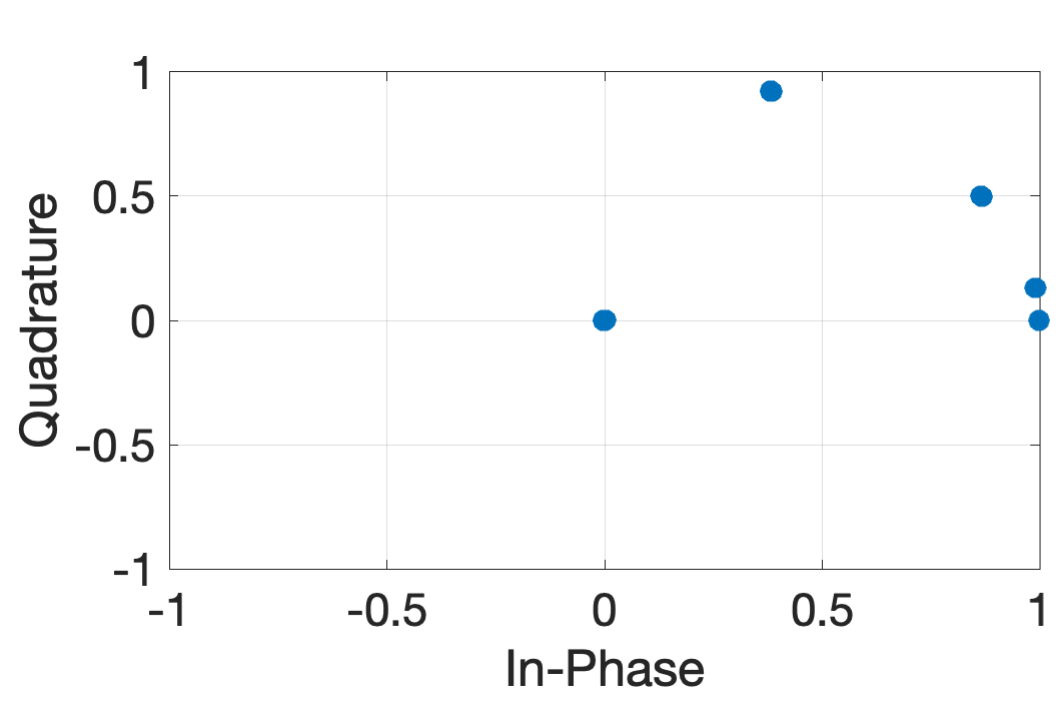}
        \caption{Radar constellation}
        \label{fig:radarconst}
    \end{subfigure}
    \caption{Radar characterization with \acrshort{psd} and constellation plots.}
    \label{fig:radarchar}
    \vspace{-5pt}
\end{figure}

In this work, we leverage a weather radar that combines techniques typical of continuous-wave radars, e.g., pulse-timing to compute the distance of the target, and of pulse radars, like the Doppler effect of the returned signal to establish the velocity of the moving target~\cite{doviak2006doppler}.
Note that similar considerations can be applied to any other radar or waveform type, and the radar signal considered in this work is a use-case study (without loss of generality) to showcase Colosseum capabilities.
Our radar operates in the S-Band, typically located within the $[3.0, 3.8]$\:GHz frequency range.
The signal has been synthetically generated as a collection of \gls{iq} samples and timestamps, with a sampling rate of $6$\:MS/s and 106657 sampling points for a total duration of $17.8$\:ms. Figure~\ref{fig:radarchar} shows some characterization of the radar signal.
Figure~\ref{fig:radarpsd} depicts the \gls{psd} of the radar.
%
Figure~\ref{fig:radarconst} displays the constellation diagram of the transmitted signal. We notice that the signal lies only in the first quadrant of the \gls{iq}-plane, which is typical of some radars.

\begin{figure}[ht]
    \centering
    \includegraphics[width=0.9\columnwidth]{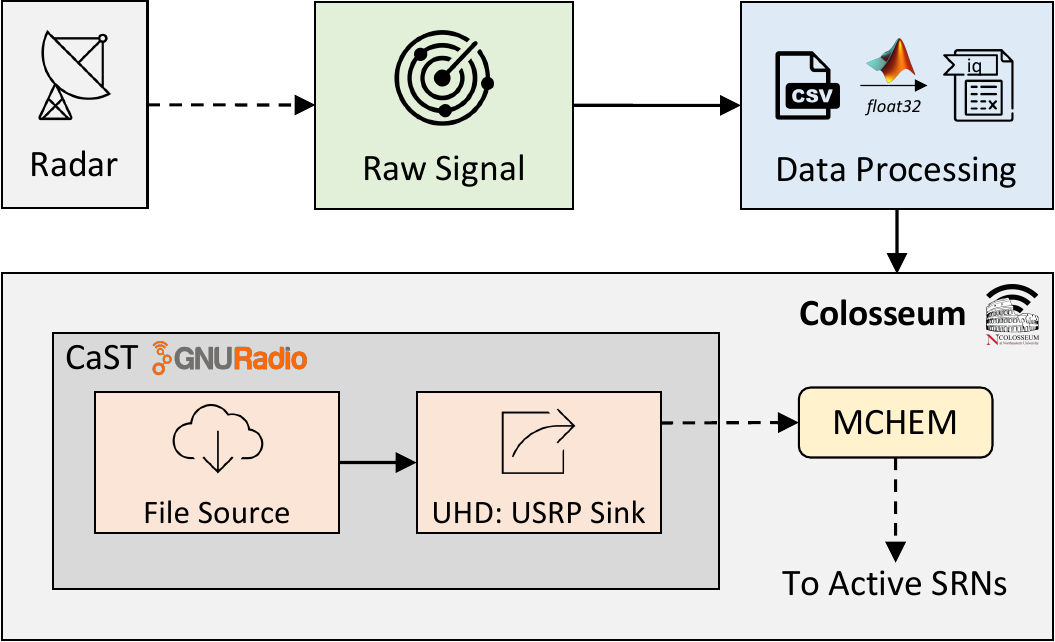}
    \caption{Block diagram of the operations needed to integrate the radar signal in Colosseum.}
    \label{fig:radarblock}
\end{figure}

Figure~\ref{fig:radarblock} shows the various operations that we developed to integrate an arbitrary waveform (radar in this case) into the Colosseum environment.
%
%
In the first step, the radar signal is generated either through a hardware device or in a synthetic manner. The output of this step is a raw signal formed of \gls{iq} samples for given time instances, which in our case is stored in a \texttt{.csv} file.
The raw data is then processed to convert it into a format that can be interpreted by Colosseum. We use MATLAB to read the \texttt{.csv} raw signal and generate a \texttt{.iq} file with the array of \gls{iq} values sequentially saved in a \texttt{float32} format.
Finally, the newly created \texttt{.iq} file is transmitted in the Colosseum environment by leveraging the open-source \gls{cast} framework, which is based on GNU Radio~\cite{villa2023dt, gnuradio}.
For the purpose of this work, \gls{cast} has been modified to include a \textit{File Source} block that allows us to load the \texttt{.iq} file on the Colosseum system. The signal is then passed through an \textit{UHD: USRP Sink} block, which connects to the USRP \gls{sdr} in Colosseum, and transmits the signal over \gls{mchem} to the other \glspl{srn}.

\begin{figure}[htp]
    \centering
    \begin{subfigure}[b]{0.49\columnwidth}
        \includegraphics[width=0.99\columnwidth]{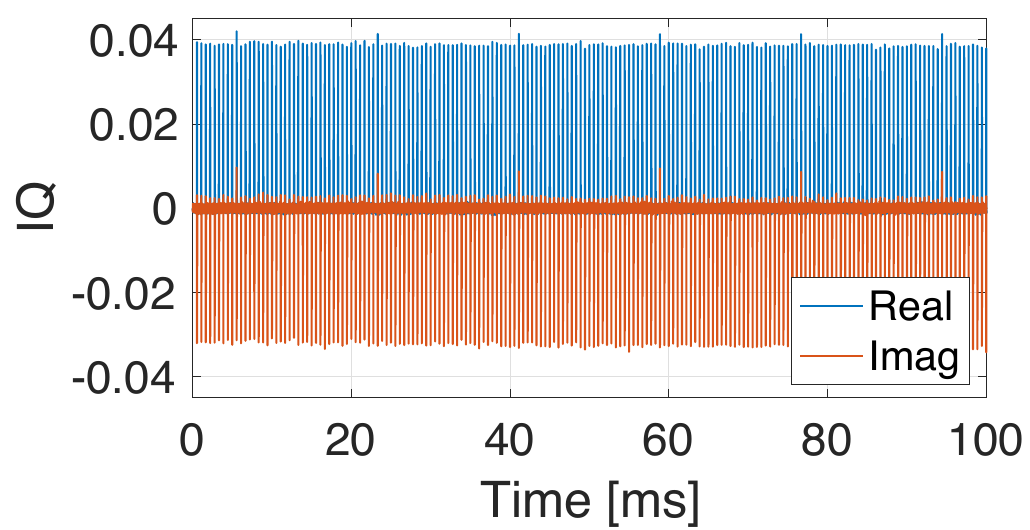}
        \caption{Received radar waveform}
        \label{fig:rxplainiq}
    \end{subfigure}
    \begin{subfigure}[b]{0.49\columnwidth}
        \includegraphics[width=0.99\linewidth]{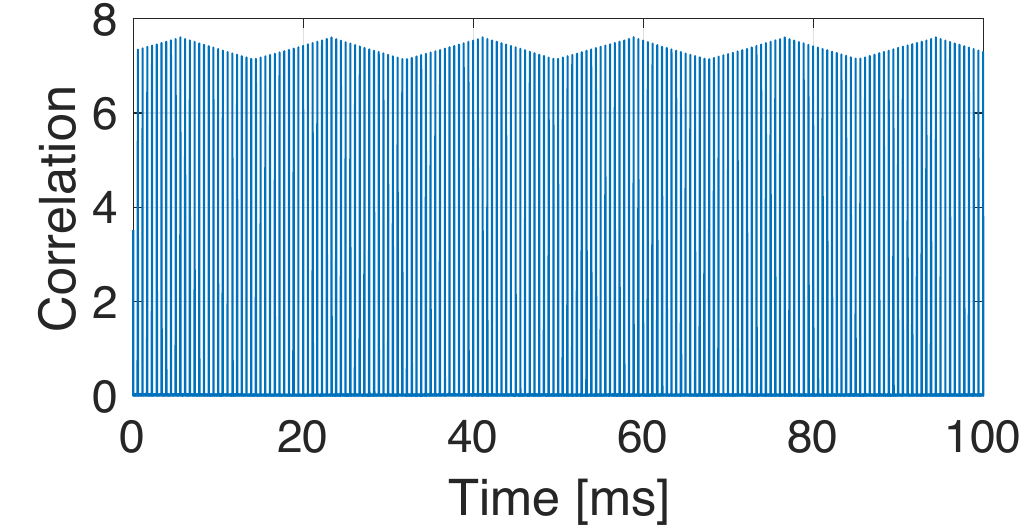}
        \caption{Radar correlation}
        \label{fig:rxcorr}
    \end{subfigure}
    \caption{Results of a radar transmission through \acrshort{cast}.}
    \label{fig:rxradarchar}
\end{figure}

%
Figure~\ref{fig:rxradarchar} shows the radar signal transmitted on the Colosseum wireless network emulator through \gls{cast} and received by another \gls{srn}.
Figure~\ref{fig:rxplainiq} displays real and imaginary parts of the raw radar waveform at the receiver node, while Figure~\ref{fig:rxcorr} the correlation between the original radar signal and the received waveform. We notice that the correlation values are clearly visible, meaning that the radar signal is correctly transmitted and detected at the receiver side.
We also notice a periodic jigsaw trend in the correlation results. This behavior is due to the large length of the transmitted radar \gls{iq} sequence (106657 complex points).
Upon performing the correlation operation, the length of the sequence causes peaks at the beginning of the sequence, as well as valleys because of leftover samples from the correlation.
%

\subsection{Colosseum Waikiki Beach Scenario}
\label{sec:rf-scenario}

\begin{figure}[t]
    \centering
    \includegraphics[width=\columnwidth]{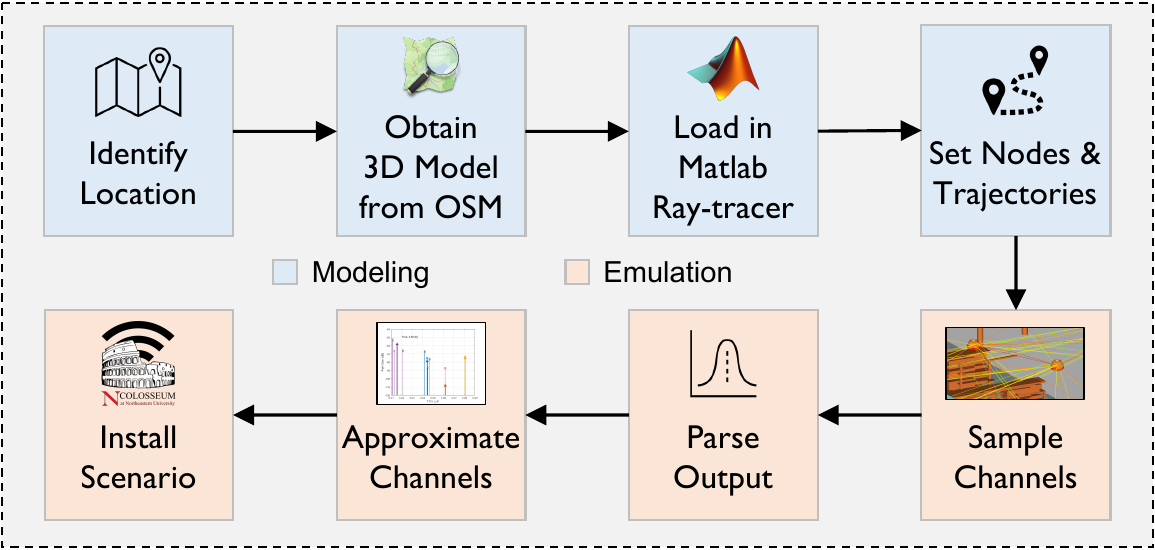}
    \caption{\gls{cast} scenario creation toolchain blocks diagram. Figure adapted from~\cite{villa2023dt}.}
    \label{fig:castscenarioblocks}
\end{figure}

To validate our use case on Colosseum, we created a novel \gls{rf} scenario that emulates the propagation environment of Waikiki Beach in Honolulu, Hawaii. This scenario involves a \gls{bs}---whose location was taken from the OpenCelliD database~\cite{opencellid} of real-world cellular deployments---that serves 6~\glspl{ue}, and a radar-equipped ship that moves in the North Pacific Ocean.
This scenario was created with the \gls{cast} toolchain~\cite{villa2023dt} following the steps of Figure~\ref{fig:castscenarioblocks}.
%

\begin{figure}[hb]
    \centering
    \includegraphics[width=0.8\columnwidth]{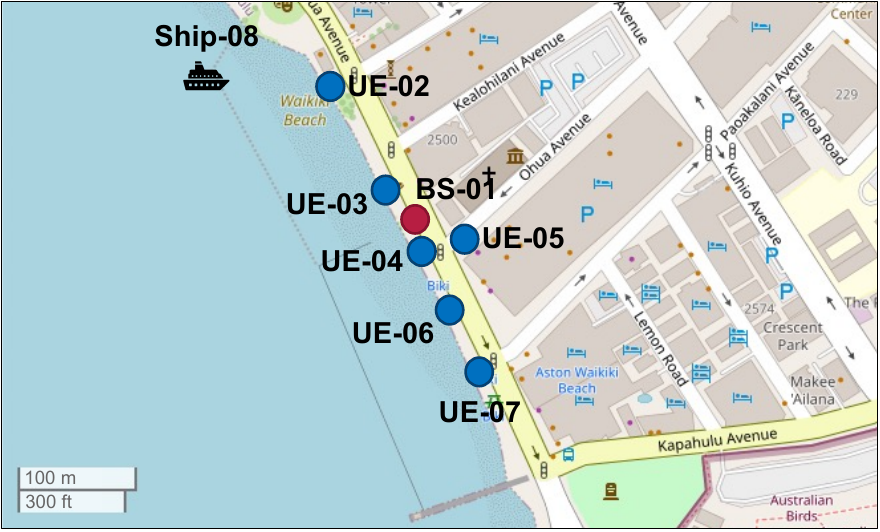}
    \caption{Location of the nodes in the Waikiki Beach scenario.}
    \label{fig:scenarionodes}
\end{figure}

In the first step, we identify the scenario location. Since we are considering a ship node for the radar, we choose the coastal area of Waikiki Beach in Honolulu, Hawaii.
Next, we obtain the 3D model of the selected location through the \gls{osm} tool~\cite{openstreetmap}. We generate an \texttt{.osm} file of a rectangular area of about $700 \times 800\:\mathrm{m}^2$, which includes Waikiki Beach, nearby buildings, and skyscrapers, as well as a portion of the ocean.
We then load the 3D model into the MATLAB ray-tracer,
and define the nodes of our scenario (shown in Figure~\ref{fig:scenarionodes}) as well as their trajectories.

\noindent
Our nodes are as follows.
\begin{itemize}[leftmargin=*]
    \item One cellular \gls{bs} (red circle in Figure~\ref{fig:scenarionodes}), whose antennas are located at $3$\:m from the ground.

    \item Six static \glspl{ue} (blue circles in Figure~\ref{fig:scenarionodes}) uniformly distributed in the surroundings of the \gls{bs}. \glspl{ue} are located at $1$\:m from the ground level to emulate hand-held devices.

    \item One ship (shown in black in Figure~\ref{fig:scenarionodes}) equipped with a radar, whose antennas are located at a height of $3$\:m.
    The ship moves following a North-South linear trajectory along Waikiki beach at a constant speed of $20$\:knots ($\sim\!\!10$\:m/s).
    This speed was derived as the average between the speed typical of civilian container ships---which travel at around $10$\:knots ($\sim\!\!5$\:m/s)---and that of aircraft carriers---which reach speeds of around $30$\:knots ($\sim\!\!15$\:m/s)~\cite{taylor2013speed}.
    
\end{itemize}

Table~\ref{table:scenarioparameters} summarizes the wireless parameters defined for the designed Colosseum \gls{rf} emulation scenario.
%
\begin{table}[htbp]
\setlength\abovecaptionskip{3pt}
    \centering
    \footnotesize
    \caption{Parameters of the Waikiki Beach scenario.}
    \label{table:scenarioparameters}
    \begin{tabularx}{0.8\columnwidth}{
        >{\raggedright\arraybackslash\hsize=1.4\hsize}X
        >{\raggedright\arraybackslash\hsize=0.6\hsize}X }
        \toprule
        Parameter & Value \\
        \midrule
        Signal bandwidth & $20$\:MHz \\
        Transmit power (BS and ship) & $30$\:dBm \\
        Transmit power (\glspl{ue}) & $20$\:dBm \\
        Antenna height (BS and ship) & $3$\:m \\
        Antenna height (\glspl{ue}) & $1$\:m \\
        Building material & Concrete \\
        Max number of reflections & $3$ \\
        Sampling time & $1$\:second \\ 
        Ship speed & $10$\:m/s \\
        Emulation area & $700x800\:\mathrm{m}^2$ \\
        \bottomrule
    \end{tabularx}
\end{table}
%
%

\begin{figure}[hb]
    \centering
    \includegraphics[width=0.8\columnwidth]{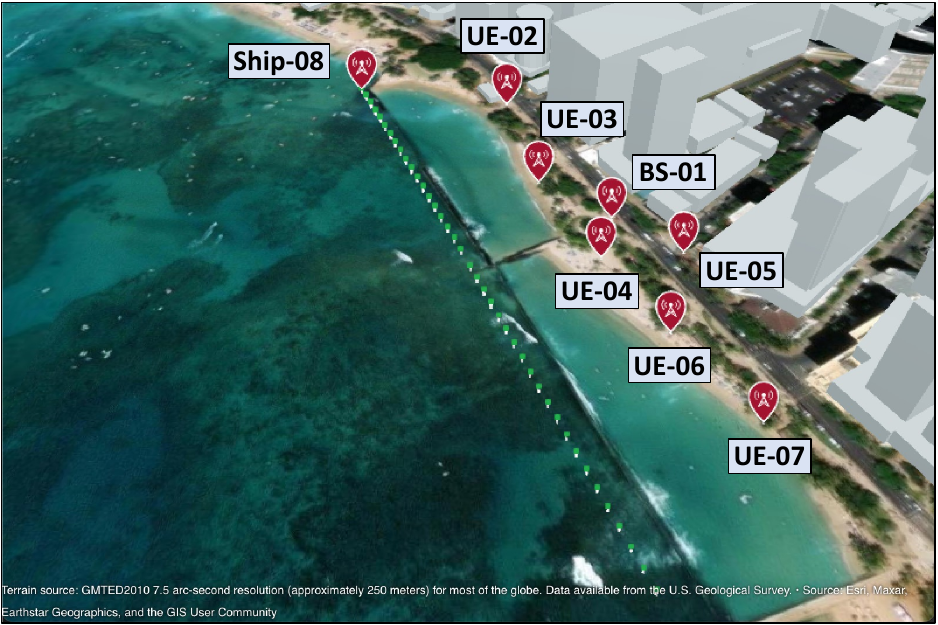}
    \caption{Layout of the scenario loaded in the MATLAB ray-tracer and visualized with Site Viewer.}
    \label{fig:scenariositeviewer}
\end{figure}

Figure~\ref{fig:scenariositeviewer} shows the layout of the scenario loaded in the MATLAB ray-tracer. We notice the 3D model of the environment (white building blocks in the figure), together with the radio node locations (red icons), and the trajectory of the ship (green dots). In this step, we perform ray-tracing to characterize the environment of interest and derive the channel taps among each pair of the nodes of our scenario.

\begin{figure}[ht]
    \centering
    \includegraphics[width=\columnwidth]{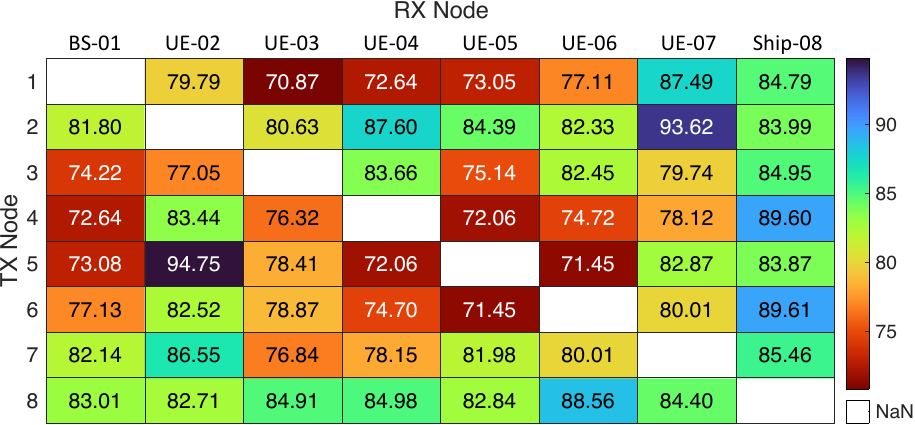}
    \caption{Heat map of the path loss among the nodes of the Waikiki scenario in Figure~\ref{fig:scenariositeviewer}. The mobile ship node is considered in its starting position on the top.}
    \label{fig:scenarioheatmap}
\end{figure}

\noindent After these operations are completed, the next step involves approximating the channel taps returned by the ray-tracer. This step is required to install the scenario in Colosseum, since \gls{mchem} supports a maximum of 4 non-zero channel taps, with a maximum delay spread
of $5.12\:\mu\mathrm{s}$.
This is performed through a k-mean clustering algorithm that we previously developed~\cite{tehrani2021creating}.
The heat map of the path loss among each pair of nodes after this channel approximation step is depicted in Figure~\ref{fig:scenarioheatmap}. (The ship node is considered to be in the top position at the beginning of the scenario.)
As expected, closer nodes experience lower path loss values.
%


As a final step, the channel taps are converted in \gls{fpga}-readable format, and the scenario is installed in Colosseum.
%
Generating the channel taps with the ray-tracing software, and approximating them to the 4 non-zero taps
required around $7$\:hours by using a 2021 MacBook Pro M1 with 10 cores and $16$\:GB of RAM.
Installing the scenario on the Colosseum system required around $50$\:minutes by leveraging a virtual machine hosted on a Dell PowerEdge M630 Server with 24 CPU cores and $96$\:GB of RAM.

\section{Intelligent Radar Detection}\label{sec:radardetection}

The \gls{bs} leverages an \gls{ai} model
to detect radar signals during or before cellular communications. This section explains how we collect and pre-process the data before feeding it into our model, as well as the structure of the model itself.

\subsection{Data Collection}

By using the scenario of Section~\ref{sec:rf-scenario}, radar and cellular signals are transmitted in different combinations and varying reception gain. Specifically, we collect \gls{iq} samples when only the radar is present, only the cellular signal is present, both are present, and neither is present (empty channel). These combinations encompass all the possible real scenarios that the intelligent radar detector might come across.

We pre-process these recordings by first breaking them into smaller samples of 1024 \glspl{iq}, as this is the input size to the \gls{ml} agent. This input size was chosen as we have found it to be the smallest size that still offers high classification performance. We then convert each sample to its frequency domain representation. Finally, we offer a binary label to each sample:~\texttt{1} if radar exists in the sample, and \texttt{0} otherwise. In this way, the model groups empty channels and un-interfered cellular transmissions as \texttt{0} and therefore be free to communicate in the given band.

\subsection{Model Design and Training}

We utilize an altered and lightweight \gls{cnn} for the radar detection. Specifically, we use a smaller version of VGG16 \cite{simonyan2014very}. We chose this structure as it is commonly used in wireless applications and can adequately show the capabilities of our framework.\footnote{More complex \gls{ai} algorithms can be used for this task. However, this is out of the scope of this paper.} 
\begin{figure}[htb]
\setlength\abovecaptionskip{1pt}
\setlength\belowcaptionskip{-5pt}
    \centering
    \includegraphics[width=\columnwidth]{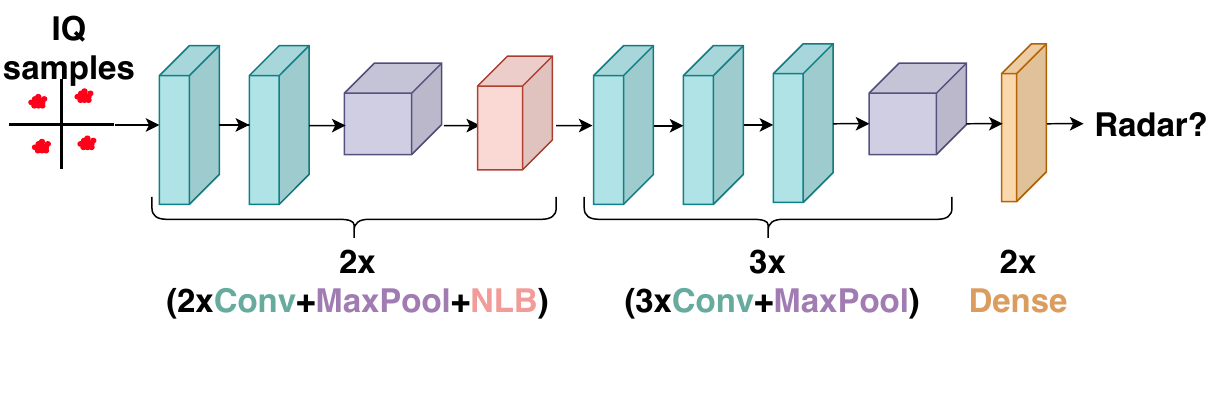}
    \vspace{-0.9cm}
    \caption{\acrshort{cnn} model used to train the radar detector.}
    \label{fig:cnn}
\end{figure}
The model architecture can be seen in Figure~\ref{fig:cnn}.
For the first two convolutional blocks, we append a \gls{nlb}. 
These blocks help the \gls{cnn} achieve self-attention and focus on spatially distant information.
More on these blocks can be read in \cite{wang2018non}.
This is a desirable trait as it can help identify long-range dependencies that may be present in the wireless signal rather than only focusing on adjacent \glspl{iq}.
The model takes as input \glspl{iq} in the shape of $(batch\_size, 1024, 2)$ where the last dimension is the real and complex part of the \gls{iq} separated into two distinct channels.

\section{Performance Evaluation}
\label{sec:results}


In this section, we present
results on
the effect of a radar signal on a cellular network deployed on the Colosseum wireless network emulator by showing: (i) the performance and accuracy of the \gls{ml} intelligent detector model; and (ii) the \glspl{kpi}, e.g., throughput, \gls{cqi}, and computation time of real-time experiments
with and without radar transmissions and the intelligent detector.

\subsection{Intelligent Detector Results}


We test our radar detector on data withheld during training and observe an accuracy of 88\%, a precision of 94\%, and a recall of 79\%. This tells us that our model is not susceptible to false positives (misclassifying empty channels or cellular signals as radar), but is susceptible to false negatives (misclassifying radar as an empty channel or a cellular signal). 

\begin{figure}[htbp]
    \centering
    \begin{subfigure}[b]{0.49\columnwidth}
        \includegraphics[width=\columnwidth]{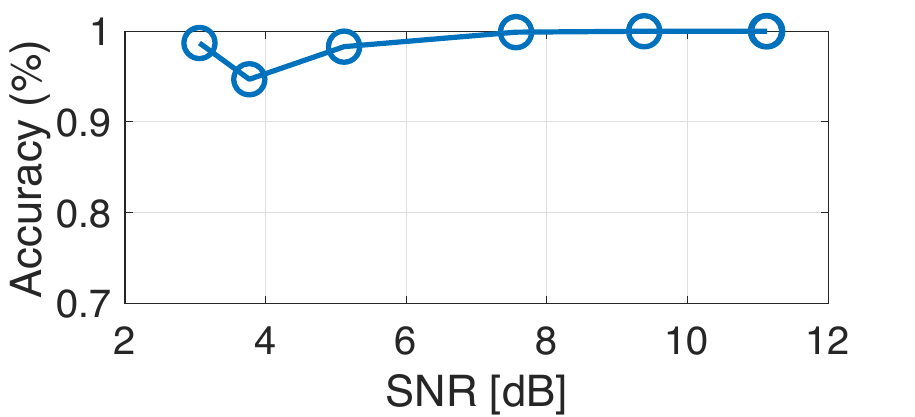}
        \caption{SNR accuracy}
        \label{fig:snracc}
    \end{subfigure}
       \hfill
    \begin{subfigure}[b]{0.49\columnwidth}
        \includegraphics[width=\linewidth]{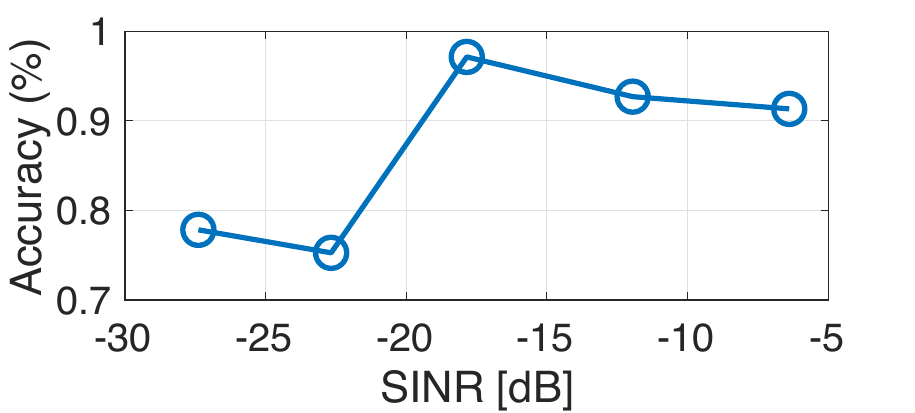}
        \caption{SINR accuracy}
        \label{fig:sinracc}
    \end{subfigure}
    \caption{\gls{cnn} radar detection accuracy with varying 
    \gls{snr} of the radar signal, and varying \gls{sinr} where the cellular signal is considered to be interference.}
    \label{fig:snrsinracc}
\end{figure}

To delve deeper into these results, we plot the accuracy as a function of \gls{snr} in Figure~\ref{fig:snracc}. For this plot, we keep the cellular nodes static in the scenario and only vary the radar gain. Here we can see that through varying \glspl{snr} we have very high and consistent performance in detecting radar signals, above 90\%. However, when we add cellular signals into the channel, this makes the detection of radar more difficult, as can be seen in Figure~\ref{fig:sinracc}, where we plot accuracy as a function of \gls{sinr}. The cellular signal is considered as interference in this plot. Indeed, with high cellular signal gain and low radar gain, we see that our classification accuracy decreases to about 75\%.  

\subsection{Experimental Results}

To properly study the network performance with and without the presence of radar transmissions, we leverage Colosseum and the newly created scenario described in Section~\ref{sec:rf-scenario} to deploy a cellular network and run traffic analysis.
The parameters of the experiments are summarized
\begin{table}[hbp]
\setlength\abovecaptionskip{3pt}
    \centering
    \footnotesize
    \caption{Parameters of the experiments.}
    \label{table:emulationparameters}
    \begin{tabularx}{0.8\columnwidth}{
        >{\raggedright\arraybackslash\hsize=1.4\hsize}X
        >{\raggedright\arraybackslash\hsize=0.6\hsize}X }
        \toprule
        Parameter & Value \\
        \midrule
        Center frequency & $3.6$\:GHz \\
        Signal bandwidth (radar) & $20$\:MHz \\
        Signal bandwidth (cellular) & $10$\:MHz \\
        Number of \glspl{bs} & 1 \\
        Number of \glspl{ue} & 6 \\
        USRP BS gains (Tx and Rx) & $[10, 30]$\:dB \\
        USRP \gls{ue} gains (Tx and Rx) & $20$\:dB \\
        USRP radar Tx gain & $20$\:dB \\
        Scenario Duration & $40$\:s \\
        Traffic type & UDP Downlink \\
        Traffic rate & $10$\:Mbps \\
        Scheduling policy & Round-robin \\
        \bottomrule
    \end{tabularx}
\end{table}
in Table~\ref{table:emulationparameters}.
The center frequency is set to $3.6$\:GHz in the newly opened \gls{cbrs} band, which is also used by S-Band-type radars. It is worth noticing that, even though characterized at $3.6$\:GHz, the scenario has been installed in Colosseum at a center frequency of $1$\:GHz, at which \gls{mchem} is optimized to work.
%
%
The scenario duration is set to $40$\:s, which is the time needed by the ship to travel the planned $400$\:m trajectory at the constant speed of $10$\:m/s.
Then, the scenario repeats cyclically from the beginning indefinitely.

We leverage the open-source SCOPE framework to deploy a twinned srsRAN protocol stack with one \gls{bs} and six \glspl{ue}~\cite{bonati2021scope,gomez2016srslte}. Additionally, the radar signal is transmitted through the use of the \gls{cast} transmit node~\cite{villa2022cast}, which has been modified to support custom radio waveforms.
In the following subsections, we show the results for three main cellular network use cases: (i) no radar transmission; (ii) with radar signal interference; and (iii) with radar and intelligent detector.
In all experiments, a \gls{udp} downlink traffic of $10$\:Mbps among \gls{bs} and \glspl{ue} is generated through iPerf, a tool to benchmark the performance of \gls{ip} networks.

\subsubsection{No Radar}
\label{sec:expnoradar}

\begin{figure}[tb]
    \centering
    \begin{subfigure}[b]{\columnwidth}
        \includegraphics[width=0.99\columnwidth]{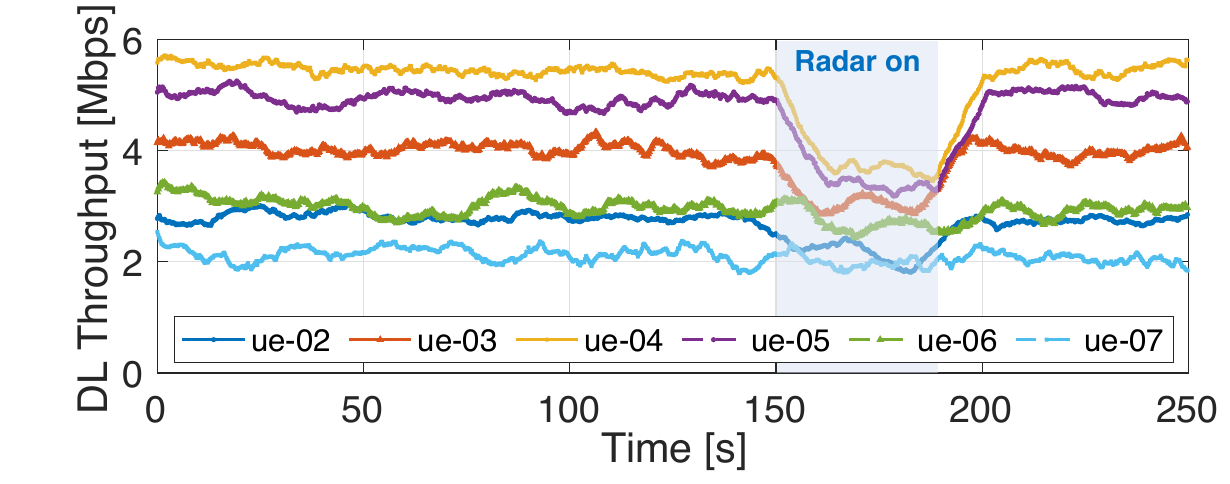}
        \caption{Downlink throughput}
        \label{fig:1resthroughput}
    \end{subfigure}
    \begin{subfigure}[b]{\columnwidth}
        \includegraphics[width=0.99\linewidth]{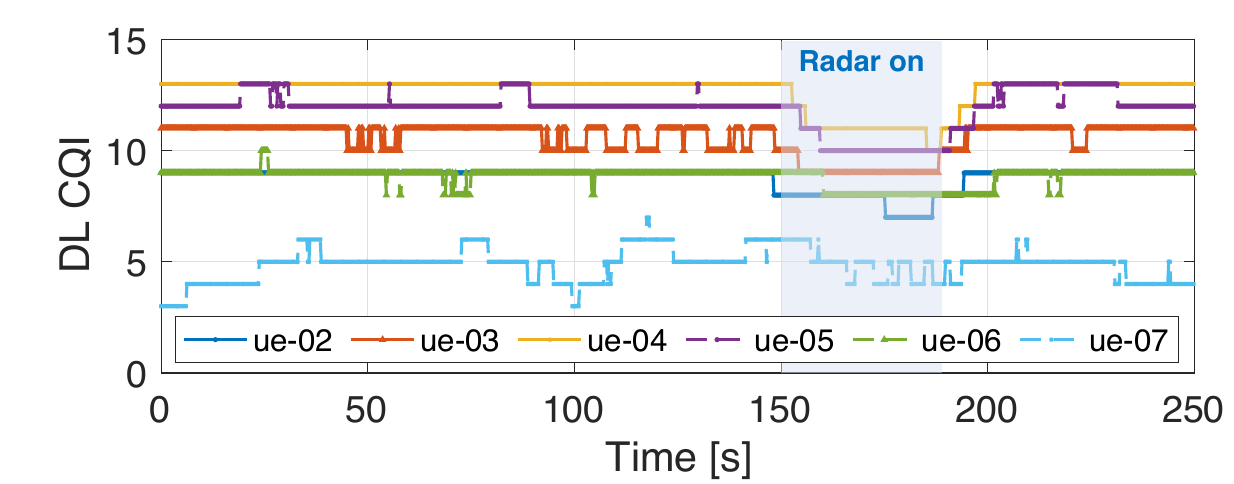}
        \caption{Downlink \gls{cqi}}
        \label{fig:1rescqi}
    \end{subfigure}
    \caption{Moving average of cellular network downlink throughput and CQI. A radar transmission is ongoing from second 150 to second 190, highlighted with a blue shade.}
    \label{fig:resultsnoradar}
    \vspace{-5pt}
\end{figure}

%
The performance of the cellular network, in terms of downlink throughput and \gls{cqi}, without radar transmissions is shown in Figure~\ref{fig:resultsnoradar}, from second $0$ to $150$.
The gains of the \gls{bs} USRP are set to $25$\:dB.%
As expected, the throughput, shown in Figure~\ref{fig:1resthroughput}, decreases with the increase of the distance between \glspl{ue} and \gls{bs}.
The best performance is achieved by UE-04, with values between $5.22$ and $5.71$\:Mbps, while the other \glspl{ue} experience a performance between $1.82$ and $5.25$\:Mbps.
The worst throughput is achieved by UE-07 due to its large distance from the \gls{bs}, environment conditions, and interference with the other nodes.
The \gls{cqi}, shown in Figure~\ref{fig:1rescqi}, follows a similar trend. Best values are reported by UE-04, with a stable \gls{cqi} of $13$.
The other \glspl{ue} show \gls{cqi} values between $2$ and $13$, with UE-07 reporting the lowest \gls{cqi} values (between $2$ and $7$).

\subsubsection{Radar}
\label{sec:exponlyradar}

The impact of radar transmissions on the cellular performance is shown in Figure~\ref{fig:resultsnoradar}, from second $150$ to $190$.
As expected, we notice a drop in the throughput (Figure~\ref{fig:1resthroughput}) and \gls{cqi} values reported by the \glspl{ue} (Figure~\ref{fig:1rescqi}).
This is more visible for the nodes closer to the \gls{bs}, e.g., UE-03, UE-04, and UE-05, since they
get more affected by the radar transmission.
When the radar stops transmitting, i.e., at around second $190$, the performance of the \glspl{ue} goes back to the initial values, i.e., to the values in the $[0, 150]$\:s window.

\subsubsection{Intelligent Radar Detection}
\label{sec:expradardetector}

\begin{figure}[t]
    \centering
    \includegraphics[width=\columnwidth]{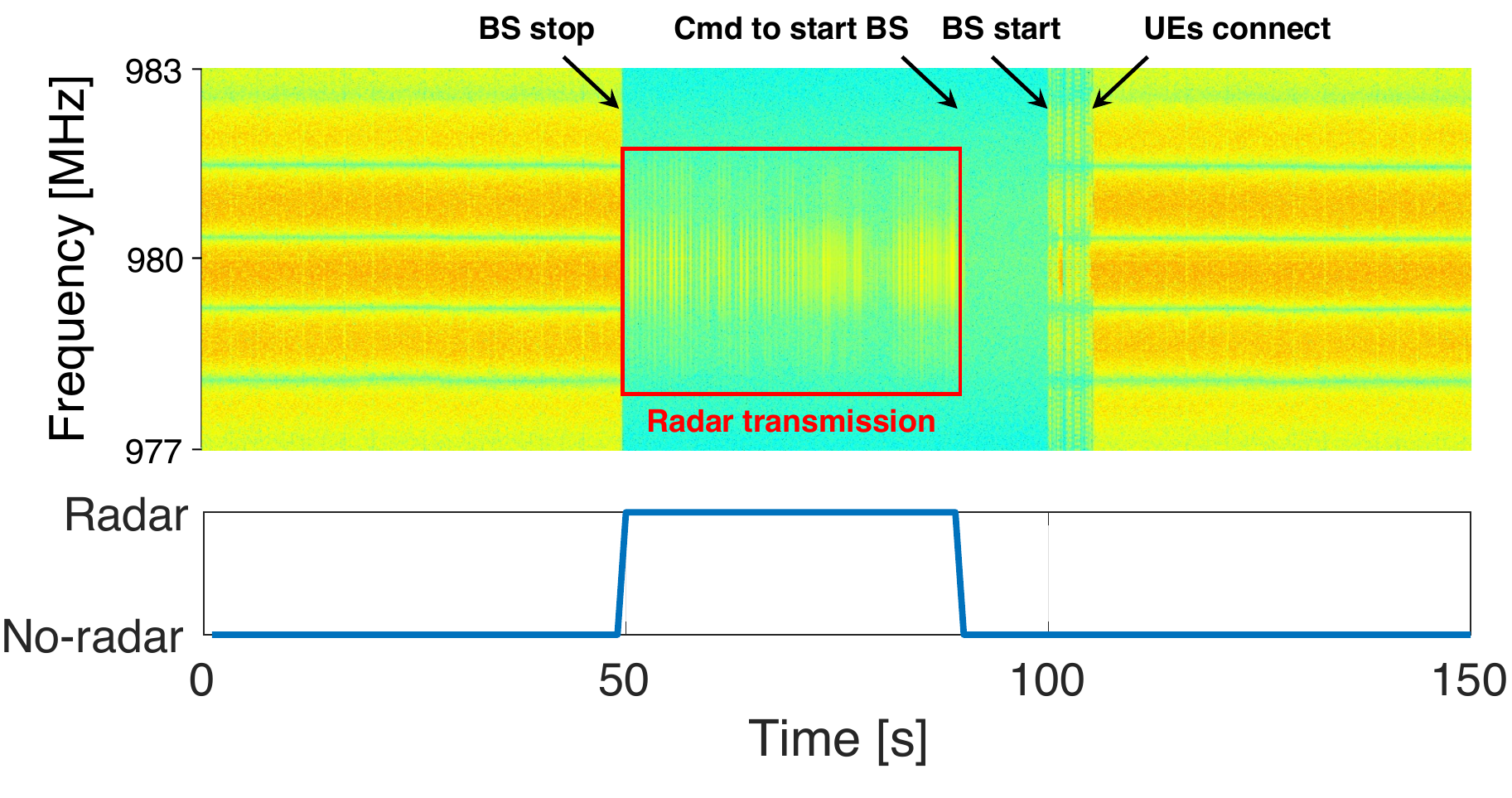}
    \caption{(top) Downlink cellular spectrogram; (bottom) radar detection system. The BS is shut down when a radar transmission is detected and resumes normal operations after no radar is detected.}
    \label{fig:expspectr}
\end{figure}

In this last use case, we evaluate the effectiveness of our intelligent detector in understanding the presence of the radar signal, as shown in Figure~\ref{fig:expspectr}.
%
The top portion of the figure shows the downlink cellular spectrogram centered at $980$\:MHz (i.e., the downlink center frequency we use for srsRAN in Colosseum) with a $6$\:MHz span; the bottom one displays the results of the radar detection system.
At the beginning of the experiment---from second $0$ to second $50$---the \gls{bs} is serving the \glspl{ue} through \gls{udp} downlink traffic (Figure~\ref{fig:expspectr}, top), as we notice from the orange and yellow stripes.
Then, at second $50$, a radar transmission is detected by the intelligent detector (Figure~\ref{fig:expspectr}, bottom) described in Section~\ref{sec:radardetection}, and the \gls{bs} is shut down accordingly.
After the radar transmission ends, i.e., at second $90$, the \gls{bs} receives the command to power back on and, after around $10$\:seconds, it resumes its operations (second $100$).
Finally, at second $110$, the \glspl{ue} reconnect to the \gls{bs}, and the downlink transmissions are restarted.
Overall, this demonstrates the effectiveness of our intelligent detector in identifying radar signals and vacating the cellular bandwidth.
Note that even if we have not tested our \gls{ml} agent with different radar signal types, changes in the radar waveform that impact its frequency domain representation might
require a re-training of the model to achieve similar performance.

\begin{figure}[b]
    \centering
    \includegraphics[width=\columnwidth]{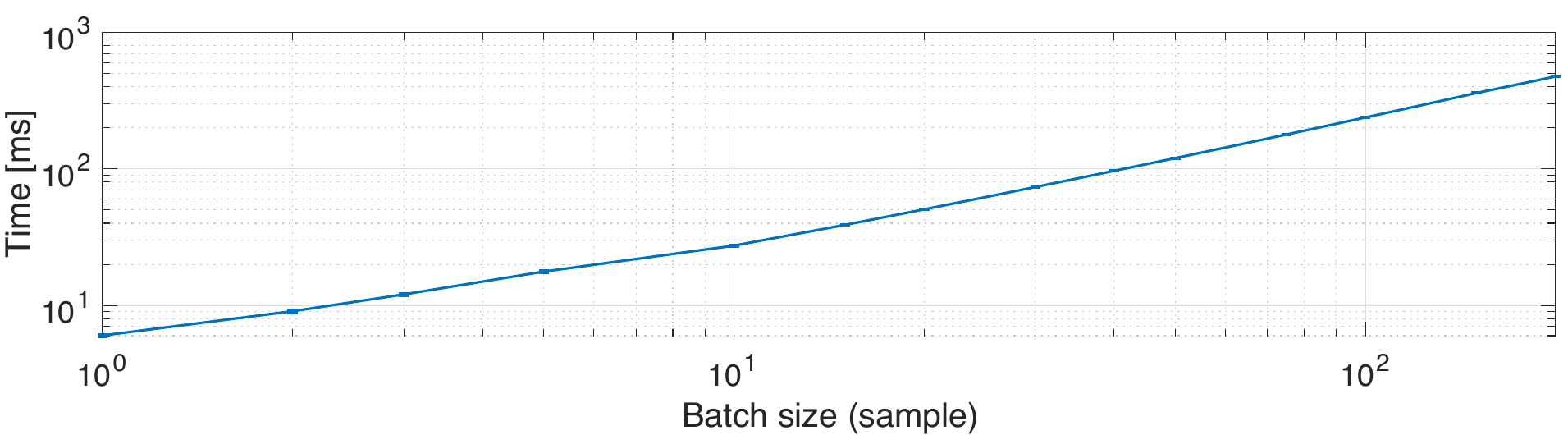}
    \caption{Computation time required for a radar classification, with different batch sizes, run on CPU on a Colosseum \gls{srn} with 48-cores Intel Xeon processor and 126\:GB of RAM.}
    \label{fig:expbatch}
\end{figure}

Figure~\ref{fig:expbatch} shows the required computation time for the classification, performed on CPU, with different batch sizes in a Colosseum \gls{srn}.
We notice that values grow linearly with the batch size, e.g., $6$\:ms for a batch size of~1 sample, $27.4$\:ms for 10, $239$\:ms for 100.
This can be traced back to the fact that these operations run on CPU, so there is not much parallelization of the processes as it would be in a GPU.
However, even in the case of a batch size of 100 samples, the maximum time of $60$\:seconds required for the detection of commercial transmissions in the \gls{cbrs} band is satisfied~\cite{goldreich2016requirements}.
To avoid false positives and false negatives, we leverage a voting system of 100 samples, in which the signal to shut down the \gls{bs} is sent only when more than $50$\% of the outputs are detecting a radar.
Moreover, we use a batch size of 10 samples, which gives us a good tradeoff between computation time and granularity of the output samples needed for the voting system.
In these conditions, our intelligent detector is able to detect an incumbent radar transmission and vacate the cellular bandwidth within $137$\:ms---which is the average time for the \gls{ml} model to generate 50 new outputs with a batch size of 10 samples---and with an accuracy of $88$\%.
%


\section{Conclusions}
\label{sec:conclusion}
In this work, we developed a framework for high-fidelity emulation-based spectrum-sharing scenarios with cellular and radar nodes implemented as a digital twin system on the Colosseum wireless network emulator.
%
First, we twinned the radar waveform on Colosseum, then we collected \gls{iq} samples of radar and cellular communications in different conditions.
Finally, we trained a \gls{cnn} network ---that can run as a dApp--- to detect the presence of the radar signal and halt the cellular network to eliminate the undesired interference on the incumbent radar communications.
Our experimental results show that our detector obtains an average accuracy of 88\% (above 90\% when \gls{snr} and \gls{sinr} are greater than $0$\:dB and $-20$\:dB respectively), and requires an average time of $137$\:ms to detect ongoing radar transmissions.

\begin{acks}
This work was partially supported by Keysight Technologies and by the U.S. National Science Foundation under grants CNS-1925601 and CNS-2112471. The views and conclusions contained in this document are those of the authors and should not be interpreted as representing the official policies, either expressed or implied, of Keysight Technologies.
\end{acks}

\balance
\bibliographystyle{ACM-Reference-Format}
\bibliography{biblio}

\end{document}

%% file: acronyms.tex
\newacronym{3gpp}{3GPP}{3rd Generation Partnership Project}
\newacronym{4g}{4G}{4th generation}
\newacronym{5g}{5G}{5th generation}
\newacronym{6g}{6G}{6th generation}
\newacronym{5gc}{5GC}{5G Core}
\newacronym{adc}{ADC}{Analog to Digital Converter}
\newacronym{aerpaw}{AERPAW}{Aerial Experimentation and Research Platform for Advanced Wireless}
\newacronym{ai}{AI}{Artificial Intelligence}
\newacronym{aimd}{AIMD}{Additive Increase Multiplicative Decrease}
\newacronym{am}{AM}{Acknowledged Mode}
\newacronym{amc}{AMC}{Adaptive Modulation and Coding}
\newacronym{amf}{AMF}{Access and Mobility Management Function}
\newacronym{aops}{AOPS}{Adaptive Order Prediction Scheduling}
\newacronym{api}{API}{Application Programming Interface}
\newacronym{apn}{APN}{Access Point Name}
\newacronym{aqm}{AQM}{Active Queue Management}
\newacronym{ausf}{AUSF}{Authentication Server Function}
\newacronym{avc}{AVC}{Advanced Video Coding}
\newacronym{awgn}{AGWN}{Additive White Gaussian Noise}
\newacronym{balia}{BALIA}{Balanced Link Adaptation Algorithm}
\newacronym{bbu}{BBU}{Base Band Unit}
\newacronym{bdp}{BDP}{Bandwidth-Delay Product}
\newacronym{ber}{BER}{Bit Error Rate}
\newacronym{bf}{BF}{Beamforming}
\newacronym{bler}{BLER}{Block Error Rate}
\newacronym{brr}{BRR}{Bayesian Ridge Regressor}
\newacronym{bsr}{BSR}{Buffer Status Report}
\newacronym{bs}{BS}{Base Station}
\newacronym{bpsk}{BPSK}{Binary Phase-shift keying}
\newacronym{bss}{BSS}{Business Support System}
\newacronym{ca}{CA}{Carrier Aggregation}
\newacronym{caas}{CaaS}{Connectivity-as-a-Service}
\newacronym{cb}{CB}{Code Block}
\newacronym{cbrs}{CBRS}{Citizens Broadband Radio Service}
\newacronym{cc}{CC}{Congestion Control}
\newacronym{ccid}{CCID}{Congestion Control ID}
\newacronym{cco}{CC}{Carrier Component}
\newacronym{cdd}{CDD}{Cyclic Delay Diversity}
\newacronym{cdf}{CDF}{Cumulative Distribution Function}
\newacronym{cdn}{CDN}{Content Distribution Network}
\newacronym{cfr}{CFR}{Code of Federal Regulations}
\newacronym{cir}{CIR}{Channel Impulse Response}
\newacronym{cn}{CN}{Core Network}
\newacronym{cnn}{CNN}{Convolutional Neural Network}
\newacronym{codel}{CoDel}{Controlled Delay Management}
\newacronym{comac}{COMAC}{Converged Multi-Access and Core}
\newacronym{cord}{CORD}{Central Office Re-architected as a Datacenter}
\newacronym{cornet}{CORNET}{COgnitive Radio NETwork}
\newacronym{cosmos}{COSMOS}{Cloud Enhanced Open Software Defined Mobile Wireless Testbed for City-Scale Deployment}
\newacronym{cots}{COTS}{Commercial Off-the-Shelf}
\newacronym{cp}{CP}{Control Plane}
\newacronym{cpu}{CPU}{Central Processing Unit}
\newacronym{cqi}{CQI}{Channel Quality Information}
\newacronym{cr}{CR}{Cognitive Radio}
\newacronym{cran}{CRAN}{Cloud \gls{ran}}
\newacronym{crs}{CRS}{Cell Reference Signal}
\newacronym{csi}{CSI}{Channel State Information}
\newacronym{csirs}{CSI-RS}{Channel State Information - Reference Signal}
\newacronym{cu}{CU}{Central Unit}
\newacronym{d2tcp}{D$^2$TCP}{Deadline-aware Data center TCP}
\newacronym{d3}{D$^3$}{Deadline-Driven Delivery}
\newacronym{dac}{DAC}{Digital to Analog Converter}
\newacronym{dag}{DAG}{Directed Acyclic Graph}
\newacronym{darpa}{DARPA}{Defense Advanced Research Projects Agency}
\newacronym{das}{DAS}{Distributed Antenna System}
\newacronym{dash}{DASH}{Dynamic Adaptive Streaming over HTTP}
\newacronym{dc}{DC}{Dual Connectivity}
\newacronym{dccp}{DCCP}{Datagram Congestion Control Protocol}
\newacronym{dce}{DCE}{Direct Code Execution}
\newacronym{dci}{DCI}{Downlink Control Information}
\newacronym{dcl}{DCL}{Dear Colleague Letter}
\newacronym{dctcp}{DCTCP}{Data Center TCP}
\newacronym{dl}{DL}{Downlink}
\newacronym{dmr}{DMR}{Deadline Miss Ratio}
\newacronym{dmrs}{DMRS}{DeModulation Reference Signal}
\newacronym{dod}{DoD}{Department of Defense}
\newacronym{drlcc}{DRL-CC}{Deep Reinforcement Learning Congestion Control}
\newacronym{drs}{DRS}{Discovery Reference Signal}
\newacronym{du}{DU}{Distributed Unit}
\newacronym{e2e}{E2E}{end-to-end}
\newacronym{ecaas}{ECaaS}{Edge-Cloud-as-a-Service}
\newacronym{ecn}{ECN}{Explicit Congestion Notification}
\newacronym{edf}{EDF}{Earliest Deadline First}
\newacronym{em}{EM}{Electromagnetic}
\newacronym{embb}{eMBB}{Enhanced Mobile Broadband}
\newacronym{empower}{EMPOWER}{EMpowering transatlantic PlatfOrms for advanced WirEless Research}
\newacronym{enb}{eNB}{evolved Node Base}
\newacronym{endc}{EN-DC}{E-UTRAN-\gls{nr} \gls{dc}}
\newacronym{epc}{EPC}{Evolved Packet Core}
\newacronym{eps}{EPS}{Evolved Packet System}
\newacronym{es}{ES}{Edge Server}
\newacronym{etsi}{ETSI}{European Telecommunications Standards Institute}
\newacronym[firstplural=Estimated Times of Arrival (ETAs)]{eta}{ETA}{Estimated Time of Arrival}
\newacronym{eutran}{E-UTRAN}{Evolved Universal Terrestrial Access Network}
\newacronym{faas}{FaaS}{Function-as-a-Service}
\newacronym{fapi}{FAPI}{Functional Application Platform Interface}
\newacronym{fcc}{FCC}{Federal Communications Commission}
\newacronym{fdd}{FDD}{Frequency Division Duplexing}
\newacronym{fdm}{FDM}{Frequency Division Multiplexing}
\newacronym{fdma}{FDMA}{Frequency Division Multiple Access}
\newacronym{fed4fire}{FED4FIRE+}{Federation 4 Future Internet Research and Experimentation Plus}
\newacronym{fft}{FFT}{Fast Fourier Transform}
\newacronym{fir}{FIR}{Finite Impulse Response}
\newacronym{fit}{FIT}{Future \acrlong{iot}}
\newacronym{fpga}{FPGA}{Field Programmable Gate Array}
\newacronym{fr2}{FR2}{Frequency Range 2}
\newacronym{fs}{FS}{Fast Switching}
\newacronym{fscc}{FSCC}{Flow Sharing Congestion Control}
\newacronym{ftp}{FTP}{File Transfer Protocol}
\newacronym{fw}{FW}{Flow Window}
\newacronym{ga128}{Ga}{Golay Sequence type A}
\newacronym{ge}{GE}{Gaussian Elimination}
\newacronym{glfsr}{GLFSR}{Galois Linear Feedback Shift Register}
\newacronym{gnb}{gNB}{Next Generation Node Base}
\newacronym{gold}{Gold}{Gold}
\newacronym{gop}{GOP}{Group of Pictures}
\newacronym{gpr}{GPR}{Gaussian Process Regressor}
\newacronym{gpu}{GPU}{Graphics Processing Unit}
\newacronym{gtp}{GTP}{GPRS Tunneling Protocol}
\newacronym{gtpc}{GTP-C}{GPRS Tunnelling Protocol Control Plane}
\newacronym{gtpu}{GTP-U}{GPRS Tunnelling Protocol User Plane}
\newacronym{gtpv2c}{GTPv2-C}{\gls{gtp} v2 - Control}
\newacronym{gw}{GW}{Gateway}
\newacronym{harq}{HARQ}{Hybrid Automatic Repeat reQuest}
\newacronym{hetnet}{HetNet}{Heterogeneous Network}
\newacronym{hh}{HH}{Hard Handover}
\newacronym{hol}{HOL}{Head-of-Line}
\newacronym{hqf}{HQF}{Highest-quality-first}
\newacronym{hss}{HSS}{Home Subscription Server}
\newacronym{http}{HTTP}{HyperText Transfer Protocol}
\newacronym{ia}{IA}{Initial Access}
\newacronym{iab}{IAB}{Integrated Access and Backhaul}
\newacronym{ic}{IC}{Incident Command}
\newacronym{ietf}{IETF}{Internet Engineering Task Force}
\newacronym{ifw}{IFW}{Interference Free Window}
\newacronym{imsi}{IMSI}{International Mobile Subscriber Identity}
\newacronym{imt}{IMT}{International Mobile Telecommunication}
\newacronym{iot}{IoT}{Internet of Things}
\newacronym{ip}{IP}{Internet Protocol}
\newacronym{iq}{IQ}{In-phase and Quadrature}
\newacronym{itu}{ITU}{International Telecommunication Union}
\newacronym{kpi}{KPI}{Key Performance Indicator}
\newacronym{kvm}{KVM}{Kernel-based Virtual Machine}
\newacronym{los}{LOS}{Line-of-Sight}
\newacronym{ls}{LS}{Loosely Synchronised}
\newacronym{lsm}{LSM}{Link-to-System Mapping}
\newacronym{lstm}{LSTM}{Long Short Term Memory}
\newacronym{lte}{LTE}{Long Term Evolution}
\newacronym{lxc}{LXC}{Linux Container}
\newacronym{m2m}{M2M}{Machine to Machine}
\newacronym{mac}{MAC}{Medium Access Control}
\newacronym{manet}{MANET}{Mobile Ad Hoc Network}
\newacronym{mano}{MANO}{Management and Orchestration}
\newacronym{mc}{MC}{Multi-Connectivity}
\newacronym{mcc}{MCC}{Mobile Cloud Computing}
\newacronym{mchem}{MCHEM}{Massive Channel Emulator}
\newacronym{mcs}{MCS}{Modulation and Coding Scheme}
\newacronym{mec}{MEC}{Multi-access Edge Computing}
\newacronym{mec2}{MEC}{Mobile Edge Cloud}
\newacronym{mfc}{MFC}{Mobile Fog Computing}
\newacronym{mi}{MI}{Mutual Information}
\newacronym{mib}{MIB}{Master Information Block}
\newacronym{miesm}{MIESM}{Mutual Information Based Effective SINR}
\newacronym{mimo}{MIMO}{Multiple Input, Multiple Output}
\newacronym{mgen}{MGEN}{Multi-Generator}
\newacronym{ml}{ML}{Machine Learning}
\newacronym{mlr}{MLR}{Maximum-local-rate}
\newacronym[plural=\gls{mme}s,firstplural=Mobility Management Entities (MMEs)]{mme}{MME}{Mobility Management Entity}
\newacronym{mmtc}{mMTC}{Massive Machine-Type Communications}
\newacronym{mmwave}{mmWave}{millimeter wave}
\newacronym{mpdccp}{MP-DCCP}{Multipath Datagram Congestion Control Protocol}
\newacronym{mptcp}{MPTCP}{Multipath TCP}
\newacronym{mr}{MR}{Maximum Rate}
\newacronym{mrdc}{MR-DC}{Multi \gls{rat} \gls{dc}}
\newacronym{mse}{MSE}{Mean Square Error}
\newacronym{mss}{MSS}{Maximum Segment Size}
\newacronym{mt}{MT}{Mobile Termination}
\newacronym{mtd}{MTD}{Machine-Type Device}
\newacronym{mtu}{MTU}{Maximum Transmission Unit}
\newacronym{mumimo}{MU-MIMO}{Multi-user \gls{mimo}}
\newacronym{mvno}{MVNO}{Mobile Virtual Network Operator}
\newacronym{nalu}{NALU}{Network Abstraction Layer Unit}
\newacronym{nas}{NAS}{Network Attached Storage}
\newacronym{nbiot}{NB-IoT}{Narrow Band IoT}
\newacronym{nfv}{NFV}{Network Function Virtualization}
\newacronym{nfvi}{NFVI}{Network Function Virtualization Infrastructure}
\newacronym{nic}{NIC}{Network Interface Card}
\newacronym{nlb}{NLB}{Non-local Block}
\newacronym{nlos}{NLOS}{Non-Line-of-Sight}
\newacronym{now}{NOW}{Non Overlapping Window}
\newacronym{nrdz}{NRDZ}{National Radio Dynamic Zone}
\newacronym{nsf}{NSF}{National Science Foundation}
\newacronym{nsm}{NSM}{Network Service Mesh}
\newacronym[type=hidden]{nr}{NR}{New Radio}
\newacronym{nrf}{NRF}{Network Repository Function}
\newacronym{nsa}{NSA}{Non Stand Alone}
\newacronym{nse}{NSE}{Network Slicing Engine}
\newacronym{nssf}{NSSF}{Network Slice Selection Function}
\newacronym{ntp}{NTP}{Network Time Protocol}
\newacronym{o2i}{O2I}{Outdoor to Indoor}
\newacronym{oai}{OAI}{OpenAirInterface}
\newacronym{oaicn}{OAI-CN}{\gls{oai} \acrlong{cn}}
\newacronym{oairan}{OAI-RAN}{\acrlong{oai} \acrlong{ran}}
\newacronym{oam}{OAM}{Operations, Administration and Maintenance}
\newacronym[plural=\gls{obu}s,firstplural=Onboard Units (OBUs)]{obu}{OBU}{Onboard Unit}
\newacronym{ofdm}{OFDM}{Orthogonal Frequency Division Multiplexing}
\newacronym{olia}{OLIA}{Opportunistic Linked Increase Algorithm}
\newacronym{omec}{OMEC}{Open Mobile Evolved Core}
\newacronym{onap}{ONAP}{Open Network Automation Platform}
\newacronym{onf}{ONF}{Open Networking Foundation}
\newacronym{onos}{ONOS}{Open Networking Operating System}
\newacronym{oom}{OOM}{\gls{onap} Operations Manager}
\newacronym{opnfv}{OPNFV}{Open Platform for \gls{nfv}}
\newacronym[type=hidden]{oran}{O-RAN}{Open \gls{ran}}
\newacronym{orbit}{ORBIT}{Open-Access Research Testbed for Next-Generation Wireless Networks}
\newacronym{os}{OS}{Operating System}
\newacronym{osm}{OSM}{Open Street Map}
\newacronym{oss}{OSS}{Operations Support System}
\newacronym{pa}{PA}{Position-aware}
\newacronym{pase}{PASE}{Prioritization, Arbitration, and Self-adjusting Endpoints}
\newacronym{pawr}{PAWR}{Platforms for Advanced Wireless Research}
\newacronym{pbch}{PBCH}{Physical Broadcast Channel}
\newacronym{pcef}{PCEF}{Policy and Charging Enforcement Function}
\newacronym{pcfich}{PCFICH}{Physical Control Format Indicator Channel}
\newacronym{pcrf}{PCRF}{Policy and Charging Rules Function}
\newacronym{pdcch}{PDCCH}{Physical Downlink Control Channel}
\newacronym{pdcp}{PDCP}{Packet Data Convergence Protocol}
\newacronym{pdsch}{PDSCH}{Physical Downlink Shared Channel}
\newacronym{pdu}{PDU}{Packet Data Unit}
\newacronym{pdp}{PDP}{Power Delay Profile}
\newacronym{pf}{PF}{Proportional Fair}
\newacronym{pgw}{PGW}{Packet Gateway}
\newacronym{phich}{PHICH}{Physical Hybrid ARQ Indicator Channel}
\newacronym{phy}{PHY}{Physical}
\newacronym{pl}{PL}{Path Loss}
\newacronym{pmch}{PMCH}{Physical Multicast Channel}
\newacronym{pmi}{PMI}{Precoding Matrix Indicators}
\newacronym{powder}{POWDER}{Platform for Open Wireless Data-driven Experimental Research}
\newacronym{ppo}{PPO}{Proximal Policy Optimization}
\newacronym{ppp}{PPP}{Poisson Point Process}
\newacronym{prach}{PRACH}{Physical Random Access Channel}
\newacronym{prb}{PRB}{Physical Resource Block}
\newacronym{psd}{PSD}{Power Spectral Density}
\newacronym{psnr}{PSNR}{Peak Signal to Noise Ratio}
\newacronym{pss}{PSS}{Primary Synchronization Signal}
\newacronym{pucch}{PUCCH}{Physical Uplink Control Channel}
\newacronym{pusch}{PUSCH}{Physical Uplink Shared Channel}
\newacronym{qam}{QAM}{Quadrature Amplitude Modulation}
\newacronym{qci}{QCI}{\gls{qos} Class Identifier}
\newacronym{qoe}{QoE}{Quality of Experience}
\newacronym{qos}{QoS}{Quality of Service}
\newacronym{qtgui}{QT-GUI}{QT Graphical User Interface}
\newacronym{quic}{QUIC}{Quick UDP Internet Connections}
\newacronym{rach}{RACH}{Random Access Channel}
\newacronym{ran}{RAN}{Radio Access Network}
\newacronym[firstplural=Radio Access Technologies (RATs)]{rat}{RAT}{Radio Access Technology}
\newacronym{rcn}{RCN}{Research Coordination Network}
\newacronym{rec}{REC}{Radio Edge Cloud}
\newacronym{red}{RED}{Random Early Detection}
\newacronym{renew}{RENEW}{Reconfigurable Eco-system for Next-generation End-to-end Wireless}
\newacronym{rf}{RF}{Radio Frequency}
\newacronym{rfc}{RFC}{Request for Comments}
\newacronym{rfr}{RFR}{Random Forest Regressor}
\newacronym{ric}{RIC}{\gls{ran} Intelligent Controller}
\newacronym{rlc}{RLC}{Radio Link Control}
\newacronym{rlf}{RLF}{Radio Link Failure}
\newacronym{rlnc}{RLNC}{Random Linear Network Coding}
\newacronym{rmse}{RMSE}{Root Mean Squared Error}
\newacronym{rnis}{RNIS}{Radio Network Information Service}
\newacronym{rr}{RR}{Round Robin}
\newacronym{rrc}{RRC}{Radio Resource Control}
\newacronym{rrm}{RRM}{Radio Resource Management}
\newacronym{rru}{RRU}{Remote Radio Unit}
\newacronym{rs}{RS}{Remote Server}
\newacronym{rsrp}{RSRP}{Reference Signal Received Power}
\newacronym{rsrq}{RSRQ}{Reference Signal Received Quality}
\newacronym{rss}{RSS}{Received Signal Strength}
\newacronym{rssi}{RSSI}{Received Signal Strength Indicator}
\newacronym{rsu}{RSU}{Road-Side Unit}
\newacronym{rtt}{RTT}{Round Trip Time}
\newacronym{ru}{RU}{Radio Unit}
\newacronym{rw}{RW}{Receive Window}
\newacronym{rx}{RX}{Receiver}
\newacronym{s1ap}{S1AP}{S1 Application Protocol}
\newacronym{sa}{SA}{standalone}
\newacronym{sack}{SACK}{Selective Acknowledgment}
\newacronym{sap}{SAP}{Service Access Point}
\newacronym{sc2}{SC2}{Spectrum Collaboration Challenge}
\newacronym{scef}{SCEF}{Service Capability Exposure Function}
\newacronym{sch}{SCH}{Secondary Cell Handover}
\newacronym{scoot}{SCOOT}{Split Cycle Offset Optimization Technique}
\newacronym{sctp}{SCTP}{Stream Control Transmission Protocol}
\newacronym{sdap}{SDAP}{Service Data Adaptation Protocol}
\newacronym{sd}{SD}{Standard Deviation}
\newacronym{sdk}{SDK}{Software Development Kit}
\newacronym{sdm}{SDM}{Space Division Multiplexing}
\newacronym{sdma}{SDMA}{Spatial Division Multiple Access}
\newacronym{sdn}{SDN}{Software-defined Networking}
\newacronym{sdr}{SDR}{Software-defined Radio}
\newacronym{seba}{SEBA}{SDN-Enabled Broadband Access}
\newacronym{sgsn}{SGSN}{Serving GPRS Support Node}
\newacronym{sgw}{SGW}{Service Gateway}
\newacronym{si}{SI}{Study Item}
\newacronym{sib}{SIB}{Secondary Information Block}
\newacronym{sinr}{SINR}{Signal to Interference plus Noise Ratio}
\newacronym{sip}{SIP}{Session Initiation Protocol}
\newacronym{siso}{SISO}{Single Input, Single Output}
\newacronym{sla}{SLA}{Service Level Agreement}
\newacronym{sm}{SM}{Saturation Mode}
\newacronym{smf}{SMF}{Session Management Function}
\newacronym{smo}{SMO}{Service Management and Orchestration}
\newacronym{sms}{SMS}{Short Message Service}
\newacronym{smsgmsc}{SMS-GMSC}{\gls{sms}-Gateway}
\newacronym{snr}{SNR}{Signal-to-Noise-Ratio}
\newacronym{son}{SON}{Self-Organizing Network}
\newacronym{sptcp}{SPTCP}{Single Path TCP}
\newacronym{srb}{SRB}{Service Radio Bearer}
\newacronym{srn}{SRN}{Standard Radio Node}
\newacronym{srs}{SRS}{Sounding Reference Signal}
\newacronym{ss}{SS}{Synchronization Signal}
\newacronym{sss}{SSS}{Secondary Synchronization Signal}
\newacronym{st}{ST}{Spanning Tree}
\newacronym{svc}{SVC}{Scalable Video Coding}
\newacronym{tb}{TB}{Transport Block}
\newacronym{tcp}{TCP}{Transmission Control Protocol}
\newacronym{tdd}{TDD}{Time Division Duplexing}
\newacronym{tdm}{TDM}{Time Division Multiplexing}
\newacronym{tdma}{TDMA}{Time Division Multiple Access}
\newacronym{tfl}{TfL}{Transport for London}
\newacronym{tfrc}{TFRC}{TCP-Friendly Rate Control}
\newacronym{tft}{TFT}{Traffic Flow Template}
\newacronym{tgen}{TGEN}{Traffic Generator}
\newacronym{tip}{TIP}{Telecom Infra Project}
\newacronym{tm}{TM}{Transparent Mode}
\newacronym{to}{TO}{Telco Operator}
\newacronym{toa}{ToA}{Time of Arrival}
\newacronym{tr}{TR}{Technical Report}
\newacronym{trp}{TRP}{Transmitter Receiver Pair}
\newacronym{ts}{TS}{Technical Specification}
\newacronym{tti}{TTI}{Transmission Time Interval}
\newacronym{ttt}{TTT}{Time-to-Trigger}
\newacronym{tx}{TX}{Transmitter}
\newacronym{uas}{UAS}{Unmanned Aerial System}
\newacronym{uav}{UAV}{Unmanned Aerial Vehicle}
\newacronym{udm}{UDM}{Unified Data Management}
\newacronym{udp}{UDP}{User Datagram Protocol}
\newacronym{udr}{UDR}{Unified Data Repository}
\newacronym{ue}{UE}{User Equipment}
\newacronym{uhd}{UHD}{\gls{usrp} Hardware Driver}
\newacronym{ul}{UL}{Uplink}
\newacronym{um}{UM}{Unacknowledged Mode}
\newacronym{uml}{UML}{Unified Modeling Language}
\newacronym{upa}{UPA}{Uniform Planar Array}
\newacronym{upf}{UPF}{User Plane Function}
\newacronym{urllc}{URLLC}{Ultra Reliable and Low Latency Communication}
\newacronym{usa}{U.S.}{United States}
\newacronym{usim}{USIM}{Universal Subscriber Identity Module}
\newacronym{usrp}{USRP}{Universal Software Radio Peripheral}
\newacronym{utc}{UTC}{Urban Traffic Control}
\newacronym{vim}{VIM}{Virtualization Infrastructure Manager}
\newacronym{vm}{VM}{Virtual Machine}
\newacronym{vnf}{VNF}{Virtual Network Function}
\newacronym{volte}{VoLTE}{Voice over \gls{lte}}
\newacronym{voltha}{VOLTHA}{Virtual OLT HArdware Abstraction}
\newacronym{vr}{VR}{Virtual Reality}
\newacronym{vran}{vRAN}{Virtualized \gls{ran}}
\newacronym{vss}{VSS}{Video Streaming Server}
\newacronym{wbf}{WBF}{Wired Bias Function}
\newacronym{wf}{WF}{Wired-first}
\newacronym{wi}{WI}{Wireless InSite}
\newacronym{wlan}{WLAN}{Wireless Local Area Network}
\newacronym{wsr}{WSR}{Weather Surveillance Radar}
\newacronym{pnf}{PNF}{Physical Network Function}
\newacronym{drl}{DRL}{Deep Reinforcement Learning}
\newacronym{mtc}{MTC}{Machine-type Communications}
\newacronym{v2x}{V2X}{Vehicle-to-everything}
\newacronym{cast}{CaST}{Channel emulation generator and Sounder Toolchain}